\documentclass[12pt]{iopart}
\usepackage{iopams}

\usepackage[authoryear,square]{natbib}

\usepackage{hyperref}

\usepackage{doi}

\def\newblock{}

\expandafter\let\csname equation*\endcsname\relax
\expandafter\let\csname endequation*\endcsname\relax

\usepackage{amsmath}
\usepackage{amsthm}

\usepackage{verbatim}

\usepackage{amsfonts}

\usepackage{dsfont}
\usepackage{graphicx}

\allowdisplaybreaks[1]

\begin{document}

\title[Visualization of Thomas-Wigner rotations]
{Visualization of Thomas-Wigner rotations}

\author{G. Beyerle}

\address{Caputh, Germany}
\ead{mail@gbeyerle.de}

\begin{abstract}
It is well known
that a sequence of two non-collinear
pure Lorentz transformations
(boosts)
is not a boost again,
but involves a spatial rotation,
the Wigner or Thomas-Wigner rotation.
The formation of this rotation
is visually illustrated
by moving a Born-rigid object
on a closed trajectory
in several sections.
Within each section
the boost's proper time duration
is assumed to be the same
and
the object's centre accelerates uniformly.
Born-rigidity implies that
the stern of this object
accelerates faster
than its bow.
It is shown
that at least five boosts
are required to return the object's centre
to its start position.
With these assumptions,
the Thomas-Wigner rotation angle
depends on a single parameter only,
the maximum speed reached
within each boost section.
The visualization highlights
the close relationship
between the Thomas-Wigner rotation
and the relativity of simultaneity.
Furthermore,
it is illustrated
that accelerated motion
implies the formation of an event horizon.
The event horizons
associated with the five boosts
constitute a boundary
to the rotated Born-rigid object
and ensure its finite size.
\end{abstract}

\maketitle

\section{Introduction}

In 1926 the British physicist Llewellyn~Hilleth~Thomas
(1903--1992)
resolved a discrepancy between
observed line splittings of atomic spectra
in an external magnetic field
(Zeeman effect)
and theoretical calculations at that time
\citep[see e.g.][]{tomonaga97}.
Thomas' analysis
\citep{thomas26,thomas27}
explains the observed deviations
in terms of a special relativistic effect
\citep{einstein05}.
He recognized
that a sequence of two non-collinear
pure Lorentz transformations (boosts)
cannot be expressed as
one single boost.
Rather,
two non-collinear boosts
correspond to a pure Lorentz transformation
combined with a spatial rotation.
This spatial rotation is known
as Wigner rotation or Thomas-Wigner rotation,
the corresponding rotation angle
is the Thomas-Wigner angle
\citep[see e.g.][and references therein]{wigner39,
benmenahem85,costella01,cushing67,ferraro99,fisher72,gelman90,
kennedy02,mocanu92,rhodes04,rowe84,ungar89,ungar97,
gourgoulhon13,misner73,rebilas15,steane12}.

The prevalent approach
to visualize Thomas-Wigner rotations
employs passive Lorentz transformations.
An object~$\mathcal{G}$
is simultaneously observed from two inertial reference frames,
$[1]$ and $[N]$,
and $\mathcal{G}$ is assumed to be at rest in both of them.
Frame~$[N]$ is related to $[1]$ by
$N-1$ pure Lorentz transformations
\begin{equation*}
[1] \rightarrow [2]
      \rightarrow [3]
      \rightarrow \ldots
      \rightarrow [N] \quad .
\end{equation*}
As already mentioned,
non-zero Thomas-Wigner rotations require
two non-collinear boosts.
For given non-collinear boosts
$[1] \rightarrow [2]$ and
$[2] \rightarrow [3]$
there exists
a unique third boost
$[3] \rightarrow [4]$,
such that
$\mathcal{G}$
is at rest with respect
to both, frame~$[1]$ and frame~$[4]$.
Thus,
non-zero Thomas-Wigner rotations imply~$N\ge4$.
In the following frame~$[1]$
is taken to be the laboratory frame.

Following \citet{jonsson07}
in the present paper
an alternative route
to visualize
Thomas-Wigner rotations
using active or ``physical'' boosts
is attempted.
That is, $\mathcal{G}$ is accelerated
starting from and ending at zero velocity
in the laboratory frame (frame~$[1]$).
During its journey
$\mathcal{G}$
performs several
acceleration and/or deceleration manoeuvres.
I suppose, that the mathematical complications
created by using active boosts
are outweighed
by the visual impressions of~$\mathcal{G}$
moving through the series of acceleration phases
and finally coming to rest in a rotated orientation
(see fig.~\ref{fg:trjboostobj} below).

The paper is sectioned as follows.
First, the approach is described
in general terms
and
the basic assumptions are introduced.
Second,
uniform accelerations of Born-rigid
objects are recalled.
In the following section
sequences of
uniform, non-collinear accelerations
for a given reference point
and the motions of its adjacent grid vertices
are calculated.
The last two sections
present the visualization results
and discuss their implications.
Details of the computer algebraic calculations
performed in this study
are given
in~\ref{se:appendix}.

For simplicity
length units of light-seconds, abbreviated ``ls''
(roughly 300,000~km)
are being used;
in these units
the velocity of light is $1~\text{ls}/\text{s}$.

\section{Method}
\label{se:method}

We consider the trajectory
of a square-shaped grid~$\mathcal{G}$.
The grid consists of~$M$ vertices
and is assumed to be Born-rigid,
i.e.\ the distances between
all grid points,
as observed in the
instantaneous comoving inertial frame,
also known as the momentarily comoving inertial frame,
remain constant
\citep{born09}.
$\mathcal{G}$'s central point~$R$
serves as the reference point.
$R$ is uniformly accelerated
for a predefined proper time period.
To obtain a closed trajectory
several of these sections
with constant proper acceleration,
but different boost direction
are joined together.

In $R$'s instantaneous comoving frame
the directions and magnitudes
of the vertices' proper accelerations~$\vec{\alpha}_i$
($i=1,\ldots,M$)
change simultaneously and discontinuously
at the switchover from one boost to the next.
The subscript~$i$ indicates
that these accelerations are in general not the same
for all vertex points.
In reference frames
other than the instantaneous comoving frame,
the accelerations~$\vec{\alpha}_i$ change asynchronously
and $\mathcal{G}$,
despite its Born-rigidity,
appears distorted and twisted
(see figure~\ref{fg:trjboostobj} below).
On the other hand,
$\mathcal{G}$'s Born-rigidity implies that
the motion of the reference point~$R$
determines the trajectories
of all remaining $M-1$ vertices;
it is sufficient to calculate
$R$'s trajectory
\citep{herglotz09,noether10}.
We note that
the separations between individual switchover events
are spacelike.
I.e.\ the switchover events are causally disconnected
and
each vertex has to be ``programmed'' in advance
to perform the required acceleration changes.

In the following,
$\alpha_R$ and $\Delta\tau_R$
denote
the magnitude of the proper acceleration
of $\mathcal{G}$'s reference point~$R$ and
the boost duration in terms of $R$'s proper time,
respectively.
To simplify the calculations
we impose the following four conditions
on all $N$ boosts.
\begin{enumerate}
\item
\label{as:bornrigid}
The grid $\mathcal{G}$ is Born-rigid.
\item
\label{as:posvelstartend}
At the start
and after completion of the $N$th boost
$\mathcal{G}$ is
at rest in frame~$[1]$;
$R$ returns to its start position
in frame~$[1]$.
\item
\label{as:constacctime}
$R$'s proper acceleration~$\alpha_R$
and the boost's proper duration~$\Delta\tau_R$
are the same in all $N$~sections.
\item
\label{as:geoplanar}
All boost directions and therefore all trajectories
lie within the $xy$-plane.
\end{enumerate}
Let the unit vector~$\hat{e}_{1}$
denote the direction of the first
boost in frame~$[1]$.
This first boost lasts for
a proper time~$\Delta\tau_R$,
as measured by $R$'s clock,
when $R$ attains
the final velocity $v_R \equiv \beta$
with respect to frame~$[1]$.
Frame~$[2]$ is now defined
as $R$'s comoving inertial frame
at this instant of time.
The corresponding
Lorentz matrix
transforming a four-vector
from frame~$[2]$ to frame~$[1]$
is
\begin{eqnarray}
\label{eq:lotra1to2}
\Lambda(\gamma,\hat{e}_1)
  \equiv
\renewcommand*{\arraystretch}{1.5}
\begin{pmatrix}
  \gamma \; ,
  &
  \gamma\,\beta\,\hat{e}^T_1
\\
  \gamma\,\beta\,\hat{e}_1 \; ,
  &
  \mathds{1}_{3{\times}3} + (\gamma-1)\,\hat{e}_1\cdot\hat{e}^T_1
\end{pmatrix} \; .
\end{eqnarray}
Here,
$\mathds{1}_{3{\times}3}$ is
the $3{\times}3$ unit matrix,
the superscript~$T$
denotes transposition,
the Lorentz factor is
\begin{eqnarray}
\label{eq:defgamma}
\gamma \equiv \frac{1}{\sqrt{1-\beta^2}}
\end{eqnarray}
and, in turn,
$\beta = \sqrt{\gamma^2-1} / \gamma$.
Similarly, frame~$[3]$ is
$R$'s comoving inertial frame
at the end of the second boost, etc.
In general,
the Lorentz transformation from
frame~$[i]$ to frame~$[i+1]$
is given by eqn.~\ref{eq:lotra1to2},
with $\hat{e}_{1}$ replaced by $\hat{e}_{i}$,
the direction of the $i$th boost in frame~$[i]$.

Assumption~[\ref{as:constacctime}]
implies
that the sole unknowns,
which need to be determined,
are the angles between consecutive boosts,
\begin{eqnarray}
\label{eq:defbeta}
\zeta_{i,i+1}
  &\equiv&
  \arccos
  \left( \hat{e}_i^T \cdot \hat{e}_{i+1} \right)
  \quad .
\end{eqnarray}
The proper acceleration~$\alpha_R$
and boost duration~$\Delta\tau_R$
are taken to be given parameters.
In the following
the boost angles~$\zeta$
will be replaced by the ``half-angle'' parametrization
\begin{eqnarray}
\label{eq:halfangleparm}
T \equiv \tan\left(\frac{\zeta}{2}\right)
\quad .
\end{eqnarray}
Eqn.~\ref{eq:halfangleparm}
allows us to write
expressions involving $\sin(\zeta)$ and $\cos(\zeta)$
as polynomials in $T$
since
\begin{eqnarray}
\label{eq:halfanglesincos}
\sin(\zeta)
  &=& \frac{2\,T}
           {1 + T^2}
  \\
\cos(\zeta)
  &=& \frac{1 - T^2}
           {1 + T^2}
  \quad .
  \nonumber
\end{eqnarray}
We will find
that a) no solutions exist
if the number of boosts~$N$ is four or less,
b) for $N = 5$ the solution is unique and
c) the derived boost angles~$\zeta$
depend solely on the selected value
of $\gamma = 1/\sqrt{1-\beta^2}$.
Changing
$\alpha_R$ and/or $\Delta\tau_R$
only affects the spatial and temporal scale
of $R$'s trajectory
(see below).

In section~\ref{se:sequniformacc} it is shown
that at least $N=5$ boost are necessary,
in order to satisfy
assumptions~[\ref{as:bornrigid}],
[\ref{as:posvelstartend}],
[\ref{as:constacctime}] and
[\ref{as:geoplanar}].
The derivations of
$\zeta_{i,i+1} = \zeta_{i,i+1}(\gamma)$
are simplified by noting
that the constraints~[\ref{as:posvelstartend}],
[\ref{as:constacctime}] and
[\ref{as:geoplanar}]
imply
time reversal invariance.
I.e.\ $R$'s trajectory from destination to start
followed backward in time
is a valid solution as well
and therefore
$\zeta_{i,i+1} = \zeta_{N-i,N-i+1}$
for $i=1,\ldots,N-1$.
Thus, for $N=5$
the number of unknown reduces from four to two,
$\zeta_{1,2} = \zeta_{4,5}$ and
$\zeta_{2,3} = \zeta_{3,4}$.

\section{Uniform acceleration of a Born-rigid object}
\label{se:uniformaccmot}

Consider the uniform acceleration
of the reference point~$R$.
We assume, the acceleration phase lasts
for the proper time period~$\Delta\tau_R$.
During~$\Delta\tau_R$
the reference point moves from location $\vec{r}_R(0)$
to location
\begin{eqnarray}
\label{eq:posattauA}
\vec{r}_R(\Delta\tau_R)
  &=&
  \vec{r}_R(0) +
  \frac{1}{\alpha_R} \,
  \left( \cosh(\alpha_R\,\Delta\tau_R) - 1 \right) \, \hat{e}_B
\end{eqnarray}
in the laboratory frame
with unit vector $\hat{e}_B$
denoting the boost direction
\citep[see e.g.][]{hamilton78,hobson06,misner73,rindler06,steane12,styer07}.
The coordinate time duration~${\Delta}t_R$
corresponding to the proper time duration~$\Delta\tau_R$
is
\begin{eqnarray}
\label{eq:timattauA}
{\Delta}t_R
  &=&
  \frac{1}{\alpha_R}\,\sinh(\alpha_R\,\Delta\tau_R)
\end{eqnarray}
and $R$'s final speed
in the laboratory frame is
\begin{eqnarray}
\label{eq:velattauA}
v_R
  = \tanh(\alpha_R\,\Delta\tau_R)
  = \beta
  \quad .
\end{eqnarray}
Let $G$ be another vertex point of the grid~$\mathcal{G}$
at location $\vec{r}_G(0)$
and
\begin{eqnarray}
\label{eq:defineofsb}
b \equiv
  \left(\vec{r}_G(0) - \vec{r}_R(0) \right)
  \cdot \hat{e}_B
\end{eqnarray}
the projection
of distance vector from $R$ to $G$
onto the boost direction~$\hat{e}_B$.
In the laboratory frame
the vertices~$G$ and $R$
start to accelerate simultaneously
since $\mathcal{G}$ is Born-rigid
and
analogous to eqns.~\ref{eq:posattauA},
\ref{eq:timattauA}
and \ref{eq:velattauA}
we obtain for $G$'s trajectory
\begin{eqnarray}
\label{eq:postimattauB}
\vec{r}_G(\Delta\tau_G)
  &=&
  \vec{r}_G(0) +
  \left( \frac{1}{\alpha_G} \,
    \left( \cosh(\alpha_G\,\Delta\tau_G) - 1 \right)
    + b
  \right) \, \hat{e}_B
  \\
{\Delta}t_G
  &=& \frac{1}{\alpha_G}\,
    \sinh(\alpha_G\,\Delta\tau_G)
  \nonumber
  \\
v_G
  &=& \tanh(\alpha_G\,\Delta\tau_G)
  \quad .
  \nonumber
\end{eqnarray}
We note, that the boost phase ends
simultaneously for all grid points
in $R$'s instantaneous comoving frame.
Since $\mathcal{G}$ is Born-rigid
(assumption~[\ref{as:bornrigid}]),
their speeds
at the end of the boost phase
have to be identical;
in particular, $v_G = v_R$
and thus
\begin{eqnarray}
\label{eq:alphatautimesalphatau}
\alpha_G\,\Delta\tau_G
  &=& \alpha_R\,\Delta\tau_R
  \quad .
\end{eqnarray}
Furthermore,
$\mathcal{G}$'s Born-rigidity implies
that the spatial distance
between $G$ and $R$
in $R$'s comoving rest frame
at the end of the boost phase
is the same as the distance
at the beginning.
A brief calculation leads to
\begin{eqnarray}
\label{eq:timspatialdist0}
-\frac{1}{\alpha_G} \, (\gamma-1)
  + b\,\gamma
  + \frac{1}{\alpha_R} \, (\gamma-1)
  = b
\end{eqnarray}
which simplifies to
\begin{eqnarray}
\label{eq:alphatfromalphar}
\alpha_G
  &=& \frac{1}{1 + b \, \alpha_R}
  \, \alpha_R
\end{eqnarray}
and with eqn.~\ref{eq:alphatautimesalphatau}
\begin{eqnarray}
\label{eq:tautfromtaur}
\Delta\tau_G
  &=& \left(1 + b \, \alpha_R\right)
  \, \Delta\tau_R
  \quad .
\end{eqnarray}
Eqn.~\ref{eq:alphatfromalphar} expresses
the well-known fact
that the proper accelerations
aboard a Born-rigid grid
may differ from
one vertex to the next.
More specifically,
at a location trailing the reference point~$R$
the acceleration exceeds $\alpha_R$,
vertex points leading~$R$ accelerate less than $\alpha_R$.
(In relativistic space travel
the passengers in the bow of the spaceship
suffer lower acceleration forces
than those seated in the stern.
This amenity of a more comfortable acceleration,
however,
is counterbalanced by faster ageing
of the space passengers
(eqn.~\ref{eq:tautfromtaur}).
Here it is assumed,
that relativistic spaceships
are Born-rigidly constructed.)

This position-dependent acceleration
is well-known from the Dewan-Beran-Bell
spaceship paradox
\citep{dewan59,dewan63}
and
\citep[chapter~9]{bell04}.
Two spaceships,
connected by a Born-rigid wire,
accelerate along the direction separating the two.
According to
eqn.~\ref{eq:alphatfromalphar}
the trailing ship has to accelerate faster
than the leading one.
Conversely,
if both accelerated at the same rate
in the laboratory frame,
Born-rigidity could not be maintained
and the wire connecting the two ships
would eventually break.
This well-known, but admittedly counterintuitive fact is
not a paradox in the true sense of the word
as discussed extensively in the relevant literature
\citep[see e.g.][]
{evett60,evett72,fernflores11,tartaglia03,redzic08,franklin10}.

Eqns.~\ref{eq:alphatfromalphar}
and
\ref{eq:tautfromtaur}
also imply,
that
$\alpha_G \rightarrow \infty$
and
$\Delta\tau_G \rightarrow 0$,
as the distance between a (trailing) vertex~$G$
and the reference point~$R$
approaches the critical value
\begin{eqnarray}
\label{eq:critdist}
  b^* \equiv -1/\alpha_R
  \quad .
\end{eqnarray}
Clearly, the object $\mathcal{G}$ cannot extend
beyond this boundary,
which in the following is referred to as ``event horizon''.
In section~\ref{su:eventhorizon} we will discuss
its consequences.

Finally, we note that
eqn.~\ref{eq:tautfromtaur} implies
that a set of initially synchronized clocks
mounted on a Born-rigid grid
will in general fall out of synchronization
once the grid is accelerated.
Thus, the switchover events,
which occur simultaneous in~$R$'s
instantaneous comoving frame,
are not simultaneous
with respect to the time
displayed by the vertex clocks.
Since switchover events
are causally not connected
and lie outside
of each others' lightcone,
the acceleration changes have to be ``programmed''
into each vertex in advance
\citep{eriksen82}.

\section{Sequence of uniform accelerations}
\label{se:sequniformacc}

In the previous section $R$'s trajectory
throughout a specific acceleration phase
(eqn.~\ref{eq:postimattauB})
was discussed.
Now several of these segments
are linked together
to form a closed trajectory for $R$.
Let $\text{A}^{[k]}$
denote $R$'s start event
as observed in frame~$[k]$
and
$\text{B}^{[k]}$,
$\text{C}^{[k]}$,
etc.\
correspondingly denote
the ``switchover'' events  between
$1^\text{st}$ and $2^\text{st}$ boost,
$2^\text{nd}$ and $3^\text{rd}$ boost,
etc.,
respectively.
The bracketed superscripts
indicate the reference frame.
Frame $[1]$, i.e.\ $k=1$, is the laboratory frame,
frame $[2]$ is obtained from frame~$[1]$
using
Lorentz transformation~$\Lambda(\gamma,-\hat{e}_1)$
(eqn.~\ref{eq:lotra1to2}).
Generally,
frame~$[k+1]$ is calculated from frame~$[k]$
using the transformation matrix~$\Lambda(\gamma,-\hat{e}_k)$.

First, we determine
the smallest number of boosts
that satisfies the four assumptions
listed in section~\ref{se:method}.
Denoting the number of boosts by $N$,
it is self-evident
that $N \ge 3$,
since
for $N = 1$
the requirement of vanishing final velocity
cannot be met
if $v_R \ne 0$.
And for $N=2$
the requirement of vanishing final velocity
implies collinear boost directions.
With two collinear boosts, however,
the reference point~$R$
does not move along a closed trajectory.
In addition, we note,
that collinear boosts
imply vanishing
Thomas-Wigner rotation
\citep[see e.g.][]{steane12}.

\subsection{Three boosts}
\label{su:threeboosts}

Consider three boosts
of the reference point~$R$
starting from location~$\text{A}$
and returning to location~$\text{D}$
via locations $\text{B}$ and $\text{C}$.
In the laboratory frame (frame~$[1]$)
the four-position
at the destination~$\text{D}$
is given by
\begin{eqnarray}
\label{eq:threeboostpos}
\textbf{\textit{P}}_\text{D}^{[1]}
  &=& \textbf{\textit{P}}_\text{A}^{[1]}
    + \textbf{\textit{S}}_{\text{A}{\rightarrow}\text{B}}^{[1]}
  \\
  && + \Lambda\left(\gamma,-\hat{e}_1\right)  \cdot
       \, \textbf{\textit{S}}_{\text{B}{\rightarrow}\text{C}}^{[2]}
  \nonumber \\
  && + \Lambda\left(\gamma,-\hat{e}_1\right) \cdot
       \, \Lambda\left(\gamma,-\hat{e}_2\right)  \cdot
       \, \textbf{\textit{S}}_{\text{C}{\rightarrow}\text{D}}^{[3]}
  \nonumber
\end{eqnarray}
and
\begin{eqnarray}
\label{eq:threeboostvel}
\textbf{\textit{V}}_\text{D}^{[1]}
  &=& \Lambda \left(\gamma,-\hat{e}_1 \right)  \cdot
      \, \Lambda \left(\gamma,-\hat{e}_2 \right)  \cdot
      \, \Lambda \left(\gamma,-\hat{e}_3 \right)  \cdot
      \, \textbf{\textit{V}}_\text{D}^{[4]}
\end{eqnarray}
is the corresponding four-velocity.
Here, the four-vector
\begin{eqnarray}
\label{eq:threeboostdeltapos}
\textbf{\textit{S}}_{\text{A}{\rightarrow}\text{B}}^{[1]}
  &\equiv&
  \frac{1}{\alpha_R} \,
  \begin{pmatrix}
  \sinh(\alpha_R\,\Delta\tau_R)
  \\
  (\cosh(\alpha_R\,\Delta\tau_R)-1)\,\hat{e}_1
  \end{pmatrix}
  \\
&=&
  \frac{1}{\alpha_R} \,
  \begin{pmatrix}
  \gamma\,\beta
  \\
  (\gamma-1)\,\hat{e}_1
  \end{pmatrix}
  \nonumber
\end{eqnarray}
describes $R$'s trajectory in spacetime
from~$\text{A}$ to~$\text{B}$
(eqn.~\ref{eq:postimattauB}).
$\textbf{\textit{S}}_{\text{B}{\rightarrow}\text{C}}^{[2]}$
and
$\textbf{\textit{S}}_{\text{C}{\rightarrow}\text{D}}^{[3]}$
are defined correspondingly.
Assumption~[\ref{as:posvelstartend}]
implies that
\begin{eqnarray}
\label{eq:posinit3boost}
\vec{P}_\text{A}^{[1]}
  = \vec{P}_\text{D}^{[1]}
  = \vec{P}_\text{A}^{[4]}
  = \vec{P}_\text{D}^{[4]}
  = \vec{0}
\end{eqnarray}
and
\begin{eqnarray}
\label{eq:velinit3boost}
\textbf{\textit{V}}_\text{A}^{[1]}
  = \textbf{\textit{V}}_\text{D}^{[1]}
  = \textbf{\textit{V}}_\text{A}^{[4]}
  = \textbf{\textit{V}}_\text{D}^{[4]}
  =
  \begin{pmatrix}
  1
  \\
  \vec{0}
  \end{pmatrix}
  \quad .
\end{eqnarray}
Inserting eqn.~\ref{eq:lotra1to2}
into eqn.~\ref{eq:threeboostvel}
yields
\begin{eqnarray}
\label{eq:threebooststvalsol}
T_{12}
  &=& T_{23}
  = \pm\sqrt{2\,\gamma+1}
\end{eqnarray}
(see~\ref{se:appendix})
and, in turn,
using eqn.~\ref{eq:posinit3boost} we obtain
\begin{eqnarray}
\label{eq:threeboostspossol}
\vec{P}_\text{D}^{[1]}
=
  \frac{1}{(\gamma + 1)^2}
  \begin{pmatrix}
  -(\gamma - 1)^2\,(2\,\gamma + 1)^2
  \\
  -(\gamma - 1)\,(2\gamma + 1)^\frac{3}{2}\,(3\gamma + 1)
  \\
  0
  \end{pmatrix}
\overset{!}{=}
  \vec{0}
  \quad .
\end{eqnarray}
Its only solution for real-valued~$\beta$
is the trivial solution $\gamma=1$,
i.e.\ $\beta=0$.
Thus, there are no non-trivial solutions
for $N=3$ boosts,
which are consistent
with the assumptions
[\ref{as:bornrigid}] to [\ref{as:geoplanar}].

\subsection{Four boosts}
\label{su:fourboosts}

For a sequence of four boosts
time reversal symmetry implies
that
$R$'s velocity in the laboratory frame
vanishes
at event~$\text{C}$
after the second boost,
i.e.\ $\vec{\text{V}}_\text{C}^{[1]} = \vec{0}$.
However, stationarity in the laboratory frame
can only be achieved
if the first two boosts
$\text{A} \rightarrow \text{B}$
and
$\text{B} \rightarrow \text{C}$
are collinear.
In order to fulfil
assumption~[\ref{as:posvelstartend}]
the third and fourth boosts
have to be collinear
with the first (and second) boost as well.
As already noted,
a sequence of collinear boosts, however,
does not produce a non-zero Thomas-Wigner rotation.

\subsection{Five boosts}
\label{su:fiveboosts}

\begin{figure}
\includegraphics{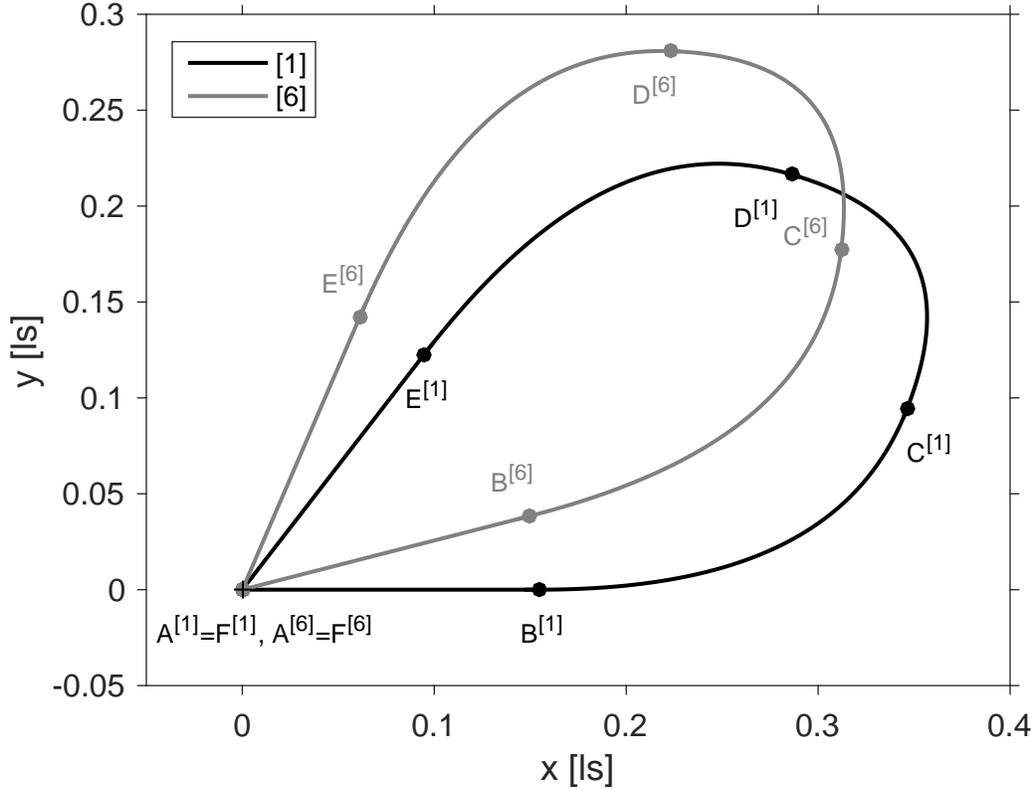}
\caption{
\label{fg:trjbstrefptfr1and6}
Trajectories of reference point~$R$
for $\gamma = 2/\sqrt{3} \approx 1.15$
as seen from reference frames
$[1]$ (laboratory frame) and $[6]$.
The two frames are stationary with respect to each
other,
but rotated by a Thomas-Wigner angle of
$\theta_{TW} = 14.4^\circ$.
}
\end{figure}

For a sequence of five boosts,
i.e.\ $N=5$,
the expressions
for four-position and four-velocity
are
\begin{eqnarray}
\label{eq:fiveboostpos}
\textbf{\textit{P}}_\text{F}^{[1]}
  &=& \textbf{\textit{P}}_\text{A}^{[1]}
    + \textbf{\textit{S}}_{\text{A}{\rightarrow}\text{B}}^{[1]}
  \\
  && + \Lambda\left(\gamma,-\hat{e}_1\right)  \cdot
       \, \textbf{\textit{S}}_{\text{B}{\rightarrow}\text{C}}^{[2]}
  \nonumber \\
  && + \Lambda\left(\gamma,-\hat{e}_1\right) \cdot
       \, \Lambda\left(\gamma,-\hat{e}_2\right)  \cdot
       \, \textbf{\textit{S}}_{\text{C}{\rightarrow}\text{D}}^{[3]}
  \nonumber \\
  && + \Lambda\left(\gamma,-\hat{e}_1\right)  \cdot
       \, \Lambda\left(\gamma,-\hat{e}_2\right)  \cdot
       \, \Lambda\left(\gamma,-\hat{e}_3\right)  \cdot
       \, \textbf{\textit{S}}_{\text{D}{\rightarrow}\text{E}}^{[4]}
  \nonumber \\
  && + \Lambda\left(\gamma,-\hat{e}_1\right)  \cdot
       \, \Lambda\left(\gamma,-\hat{e}_2\right)  \cdot
       \, \Lambda\left(\gamma,-\hat{e}_3\right)  \cdot
       \, \Lambda\left(\gamma,-\hat{e}_4\right)  \cdot
       \, \textbf{\textit{S}}_{\text{E}{\rightarrow}\text{F}}^{[5]}
  \nonumber
\end{eqnarray}
and
\begin{eqnarray}
\label{eq:fiveboostvel}
\textbf{\textit{V}}_\text{F}^{[1]}
  &=& \Lambda\left(\gamma,-\hat{e}_1\right)    \cdot
      \, \Lambda\left(\gamma,-\hat{e}_2\right) \cdot
      \, \Lambda\left(\gamma,-\hat{e}_3\right)
  \\
  &&  \cdot
      \, \Lambda\left(\gamma,-\hat{e}_4\right) \cdot
      \, \Lambda\left(\gamma,-\hat{e}_5\right) \cdot
      \, \textbf{\textit{V}}_\text{F}^{[6]}
  \quad ,
  \nonumber
\end{eqnarray}
respectively.
Analogous to eqns.~\ref{eq:posinit3boost}
and \ref{eq:velinit3boost}
assumption [\ref{as:posvelstartend}]
implies that
\begin{eqnarray}
\label{eq:posinit5boost}
\vec{P}_\text{A}^{[1]}
  = \vec{P}_\text{F}^{[1]}
  = \vec{P}_\text{A}^{[6]}
  = \vec{P}_\text{F}^{[6]}
  = \vec{0}
\end{eqnarray}
and
\begin{eqnarray}
\label{eq:velinit5boost}
\textbf{\textit{V}}_\text{A}^{[1]}
  = \textbf{\textit{V}}_\text{F}^{[1]}
  = \textbf{\textit{V}}_\text{A}^{[6]}
  = \textbf{\textit{V}}_\text{F}^{[6]}
  =
  \begin{pmatrix}
  1
  \\
  \vec{0}
  \end{pmatrix}
  \quad .
\end{eqnarray}
To simplify the expressions
in eqns.~\ref{eq:fiveboostpos} and
\ref{eq:fiveboostvel}
time reversal symmetry
is invoked again.
It implies
that the set of boost vectors
$-\hat{e}_5, -\hat{e}_4, \ldots, -\hat{e}_1$
constitutes a valid solution,
provided
$\hat{e}_1, \hat{e}_2, \ldots, \hat{e}_5$
is one and satisfies
assumptions~[\ref{as:bornrigid}] to [\ref{as:geoplanar}].
Thereby the number of unknown is reduced
from four to two,
the angle between the boost vectors
$\hat{e}_1$ and $\hat{e}_2$,
and the angle between
$\hat{e}_2$ and $\hat{e}_3$
\begin{eqnarray}
\label{eq:defbeta123}
\zeta_{1,2}
  &\equiv&
  \arccos\left( \hat{e}_1^T \cdot \hat{e}_2\right)
  =
  \arccos\left( \hat{e}_4^T \cdot \hat{e}_5\right)
  \\
\zeta_{2,3}
  &\equiv&
  \arccos\left( \hat{e}_2^T \cdot \hat{e}_3\right)
  =
  \arccos\left( \hat{e}_3^T \cdot \hat{e}_4\right)
  \quad .
  \nonumber
\end{eqnarray}
Fig.~\ref{fg:trjbstrefptfr1and6} illustrates
the sequence of the five boosts
in the laboratory frame
(frame~$[1]$).
$R$ moves along a closed trajectory
starting at location~$\text{A}$ and returning to $\text{F}$
via $\text{B}$, $\text{C}$, $\text{D}$ and $\text{E}$.
Since the start and final velocities are zero,
the motion between $\text{A}$ and $\text{B}$
and, likewise, between $\text{E}$ and $\text{F}$
is rectilineal.
In contrast,
the trajectory connecting
$\text{B}$ and $\text{E}$ (via $\text{C}$ and $\text{D}$)
appears curved in frame~$[1]$.
As discussed and illustrated below
the curved paths appear as straight lines
in the corresponding frame
(see fig.\ref{fg:trjboostrefpt}).

From eqns.~\ref{eq:fiveboostvel}
and \ref{eq:velinit5boost}
it follows that
\begin{equation}
\label{eq:t12dott23}
  \left( (T_{12})^2 - 2 \, \gamma - 1\right) \, (T_{23})^2
  - 4 \, (1+\gamma) \, T_{12} \, T_{23}  + (T_{12})^2
  + 4 \, \gamma^2 + 2 \, \gamma - 1 = 0
\end{equation}
with the two unknowns $T_{12}$ and $T_{23}$
(for details see~\ref{se:appendix}).
Eqn.~\ref{eq:t12dott23} has the two solutions
\begin{eqnarray}
\label{eq:t23fromt12}
T_{23}^{(\pm)}
  &=&
  \frac{1}{- (T_{12})^2 + 2 \gamma + 1}
  \\
&&
  \times
  \bigg(- 2\,T_{12}\,\left(\gamma + 1\right)
  \nonumber \\
&&
  \quad \pm \sqrt{- (T_{12})^4 + 8\,(T_{12})^2 \gamma
  + 6\,(T_{12})^2
  + 8\,\gamma^3 + 8\,\gamma^2 - 1} \bigg)
  \nonumber
\end{eqnarray}
provided
\begin{eqnarray}
\label{eq:t12exprneqzero}
(T_{12})^2 - 2 \gamma - 1 \ne 0
  \quad .
\end{eqnarray}
Assumption~[\ref{as:posvelstartend}] implies
that $\vec{P}^{[1]}_\text{F} = 0$,
i.e.\ the spatial component of
the event
$\textbf{\textit{P}}_\text{F}^{[1]}$
vanishes.
Since all motions are restricted to the $xy$-plane,
it suffices to consider the $x$- and $y$-components
of eqn.~\ref{eq:posinit5boost}.
The $y$-component leads to the equations
(see~\ref{se:appendix})
\begin{eqnarray}
\label{eq:t12t23first}
&& (T_{12})^{8}
\\
&&
+ (T_{12})^{6} \, 4 \,
  \left(\gamma + 2\right)
  \nonumber
  \\
&&
+ (T_{12})^4 \, (-2) \,
  \left(2\,\gamma+1\right) \, \left(2\,\gamma^2 + 8\,\gamma + 9\right)
  \nonumber
  \\
&&
+ (T_{12})^2 \, (-4) \,
  \left(8 \gamma^4 + 28 \gamma^3 + 26 \gamma^2 + 5 \gamma - 2\right)
  \nonumber
  \\
&&
- \left(2 \gamma + 1\right)^3 \, \left(4\,\gamma^2 + 2 \gamma - 1\right)
  \nonumber
  \\
&=& 0
\nonumber
\end{eqnarray}
or
\begin{eqnarray}
\label{eq:t12t23second}
&& (T_{12})^{8} \,
  \left(\gamma^2 + 6\,\gamma + 9 \right)
\\
&&
+ (T_{12})^{6} \, (-4) \,
  \left(6 \, \gamma^3 - 15\,\gamma^2 - 12\,\gamma + 5 \right)
  \nonumber
  \\
&&
+ (T_{12})^4 \, (-2) \,
  \left(24\,\gamma^4 - 44\,\gamma^3 - 55\,\gamma^2 + 26\,\gamma + 1 \right)
  \nonumber
  \\
&&
+ (T_{12})^2 \, (-4) \,
  \left( 8\,\gamma^5 - 16\,\gamma^4 - 14\,\gamma^3 + 5\,\gamma^2 + 1 \right)
  \nonumber
  \\
&&
+ \left( 4 \, \gamma^2 + \gamma - 1 \right)^2
  \nonumber
  \\
&=& 0 \quad .
\nonumber
\end{eqnarray}
For $\gamma=1$
eqn.~\ref{eq:t12t23second}
yields
\begin{eqnarray}
\label{eq:t12t23atunitgamma}
(T_{12})^8
  + 4\,(T_{12})^6
  + 6\,(T_{12})^4
  + 4\,(T_{12})^2
  + 1 = 0
\end{eqnarray}
which has no real-valued solution for $T_{12}$.

It turns out (see~\ref{se:appendix})
that equating the $x$-component of $\vec{P}^{[1]}_\text{F}=0$
to zero,
results
in an expression containing two factors as well,
one of which is identical
to the LHS of eqn.~\ref{eq:t12t23first}.
Thus, eqn.~\ref{eq:t12t23first}
satisfies the condition $\vec{P}^{[1]}_\text{F}=0$.

Eqn.~\ref{eq:t12t23first} is a polynomial equation of degree four
in terms of $(T_{12})^2$.
Its solutions are classified
according to the value of the discriminant~$\Delta$,
which for eqn.~\ref{eq:t12t23first} evaluates to
(see e.g.\ \url{en.wikipedia.org/wiki/Quartic_function})
\begin{eqnarray}
\label{eq:pol4degdelta}
\Delta &=&
  -524288 \, \gamma
          \,(\gamma - 1)^3
          \,(\gamma + 1)^7
\\
&&
  \times\,(4\,\gamma^4 + 28\,\gamma^3 + 193\,\gamma^2 + 234\,\gamma + 81)
  \quad .
  \nonumber
\end{eqnarray}
For non-trivial boost $\gamma>1$,
the discriminant is negative
and therefore
the roots of the quartic polynomial
consist of two pairs of
real and complex conjugate numbers.
The real-valued solutions are
\begin{eqnarray}
\label{eq:solvepol4deg}
(T_{12})^2
  &=& - (\gamma+2) + \mathcal{S}
      + \frac{1}{2} \,
        \sqrt{-4\,\mathcal{S}^2 - 2\,p - \frac{q}{\mathcal{S}}}
\end{eqnarray}
and
\begin{eqnarray}
\label{eq:solvepol4degD}
\left(T_{12}^{(b)}\right)^2
  &=& - (\gamma+2) + \mathcal{S}
      - \frac{1}{2} \,
        \sqrt{-4\,\mathcal{S}^2 - 2\,p - \frac{q}{\mathcal{S}}}
\end{eqnarray}
with
\begin{eqnarray}
\label{eq:pol4pars}
p
  &\equiv& -2 \, \left( \gamma + 1 \right)
         \, \left( 4\gamma^2 + 17 \gamma + 21 \right)
  \\
q
  &\equiv& -16 \, \left( \gamma + 1 \right)
          \, \left( \gamma^2 - 2 \gamma -9 \right)
  \nonumber
  \\
\mathcal{S}
  &\equiv& \frac{1}{2\,\sqrt{3}} \,
    \sqrt{ -2 \, p + \mathcal{Q} + \frac{\Delta_0}{\mathcal{Q}} }
  \nonumber
  \\
\Delta_0
  &\equiv&
  16 \, \left( \gamma + 1 \right)^2 \,
     \, \left( 4\,\gamma^4 + 28\,\gamma^3 + 157\,\gamma^2
              + 126\,\gamma + 9 \right)
  \nonumber
  \\
\mathcal{Q}
  &\equiv&
  4 \, \sqrt[3]{ \mathcal{Q}_0 + 12 \, \sqrt{6} \, \sqrt{\mathcal{Q}_0} }
  \nonumber
  \\
\mathcal{Q}_0
  &\equiv&
  \gamma
  \, ( \gamma - 1 )^3
  \, ( \gamma + 1 )^7
  \, ( 4 \, \gamma^4 + 28 \, \gamma^3 + 193 \, \gamma^2 + 234 \, \gamma + 81 )
  \quad .
  \nonumber
\end{eqnarray}
The solution
listed in eqn.~\ref{eq:solvepol4degD}
turns out to be negative
and thus does not produce a real-valued
solution for
$T_{12}$.
The remaining two solutions
of eqn.~\ref{eq:t12t23first}
\begin{eqnarray}
\label{eq:t12othersols}
\left(T_{12}^{(c)}\right)^2
  &=& - (\gamma+2) - \mathcal{S}
    + \frac{1}{2} \,
      \sqrt{-4\,\mathcal{S}^2 - 2\,p + \frac{q}{\mathcal{S}}}
  \\
\left(T_{12}^{(d)}\right)^2
  &=& - (\gamma+2) - \mathcal{S}
    - \frac{1}{2} \,
      \sqrt{-4\,\mathcal{S}^2 - 2\,p + \frac{q}{\mathcal{S}}}
  \nonumber
\end{eqnarray}
correspond to replacing
$\mathcal{S} \rightarrow -\mathcal{S}$
in eqns.~\ref{eq:solvepol4deg}
and \ref{eq:solvepol4degD},
are complex-valued
and therefore disregarded as well.

The other unknown, $T_{23}$,
follows from eqn.~\ref{eq:t23fromt12}
by choosing the positive square root
$+\sqrt{(T_{12})^2}$
and using
\begin{eqnarray}
\label{eq:t23fromt12sel}
T_{23}
  &\equiv& T_{23}^{(+)}
\end{eqnarray}
(see eqn.~\ref{eq:t12dott23}).
For a given
Lorentz factor~$\gamma$
the angles between
the boost directions~$\hat{e}_i$
and $\hat{e}_{i+1}$
follow from
\begin{eqnarray}
\label{eq:zeta1fromgamma}
\zeta_{1,2}(\gamma)
  &=& \arccos\left( \hat{e}^T_1 \cdot \hat{e}_2 \right)
  = 2 \, \arctan\left( +\sqrt{(T_{12}(\gamma))^2} \right)
  \\
\zeta_{2,3}(\gamma)
  &=& \arccos\left( \hat{e}^T_2 \cdot \hat{e}_3 \right)
  = 2 \, \arctan\left( +\sqrt{(T_{23}(\gamma))^2} \right)
  \nonumber
\end{eqnarray}
and,
with $\zeta_{4,5} = \zeta_{1,2}$
and
$\zeta_{3,4} = \zeta_{2,3}$,
the orientation of the five boost directions
$\hat{e}_i$ for $i=1,\ldots,5$
within the $xy$-plane are obtained.

\begin{figure}
\includegraphics{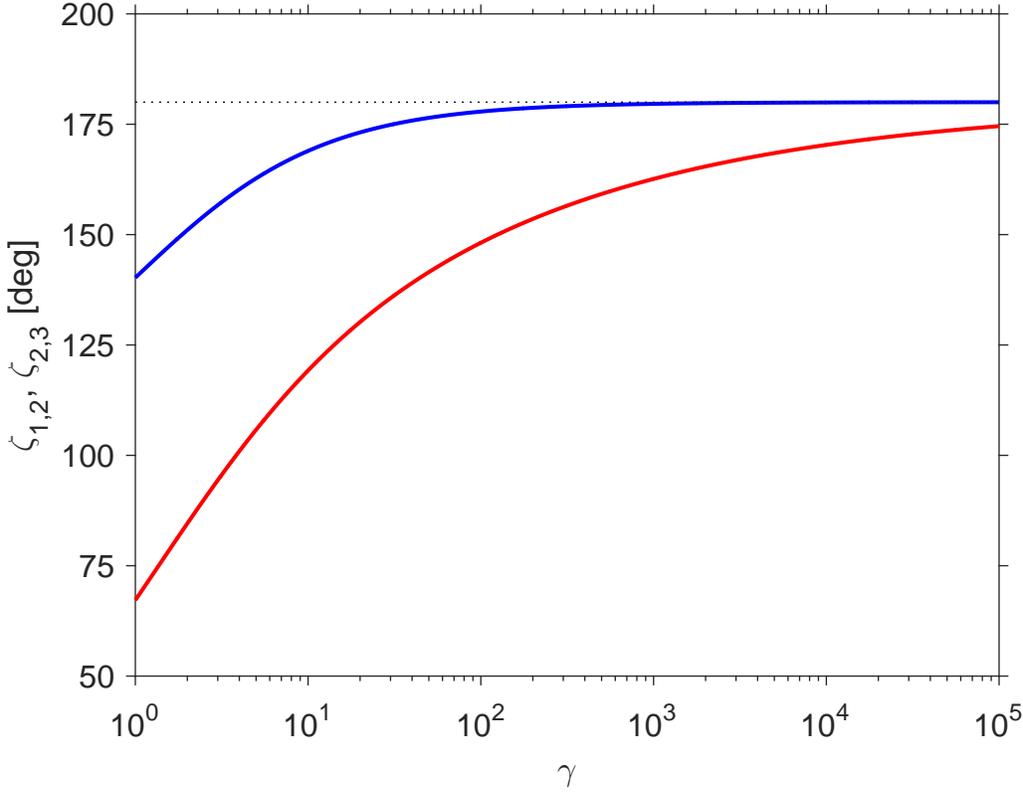}
\caption{
\label{fg:boostanglesvsgamma}
The angle between the boost direction vectors
$\hat{e}_1$ and $\hat{e}_2$ in frame~$[1]$ (blue line),
and the angle
between $\hat{e}_2$ and $\hat{e}_3$ in frame~$[2]$ (red)
as a function of $\gamma$.
The dotted line marks $+180^\circ$,
the limit of $\zeta_{1,2}$ and $\zeta_{2,3}$
for $\gamma\rightarrow\infty$.
}
\end{figure}

Fig.~\ref{fg:boostanglesvsgamma}
shows numerical values of the boost angles
$\zeta_{1,2}$ and $\zeta_{2,3}$
as a function of $\gamma$.
The angles increase
from
\[
\zeta_{1,2}(\gamma=1)
  = 2 \, \arctan\left(\sqrt{-5 + 4\,\sqrt{10}}\right)
  \approx 140.2^\circ
\]
and
\[
\zeta_{2,3}(\gamma=1)
  = 2 \, \arctan\left(
    \frac{\sqrt{-5 + 4\,\sqrt{10}}
          - \sqrt{3}\,\sqrt{-5 + 2\,\sqrt{10}}}
         {-2 + \sqrt{10}} \right)
  \approx 67.2^\circ
\]
at $\gamma=1$ to
$\zeta_{1,2}(\gamma \rightarrow \infty) = +180^\circ =
\zeta_{2,3}(\gamma \rightarrow \infty)$
as $\gamma \rightarrow \infty$.

\begin{figure}
\includegraphics{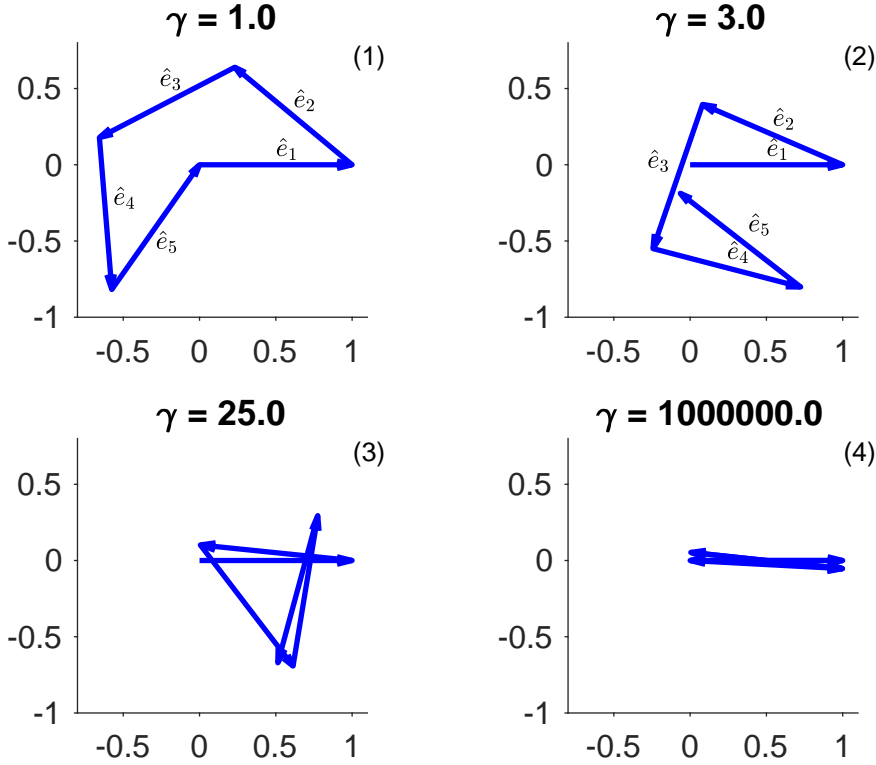}
\caption{
\label{fg:normboostdirs}
Boost directions for four different
values of $\gamma = 1 / \sqrt{1-\beta^2}$.
$\hat{e}_k$ is assumed to point along the $x$-axis.
In the relativistic limit $\gamma \rightarrow \infty$
the angles between $\hat{e}_k$ and $\hat{e}_{k+1}$
approach $+180^\circ$
and the motion of the reference point~$R$
tends to be more and more
restricted along the $x$-axis (see panel~(4)),
i.e.\ in the relativistic limit
the object's trajectory
transitions from a two- to a one-dimensional
motion.
}
\end{figure}

Fig.~\ref{fg:normboostdirs} depicts
the orientation of the five boost directions
for several values of $\gamma$.
For illustrative purposes
the first boost vector~$\hat{e}_1$
is taken to point along the $x$-axis.
We note
that the panels in fig.~\ref{fg:normboostdirs}
do not represent a specific reference frame;
rather,
each vector $\hat{e}_k$ is plotted
with respect to frame~$[k]$
($k=1,\ldots,5$).
The four panels show the changes in boost directions
for increasing values of $\gamma$.
Interestingly,
the asymptotic limits
$\zeta_{1,2}(\gamma\rightarrow\infty) = +180^\circ$
and
$\zeta_{2,3}(\gamma\rightarrow\infty) = +180^\circ$
imply
that in the relativistic limit
$\gamma\rightarrow\infty$
the trajectory of~$R$
essentially reduces
to one-dimensional motions along the $x$-axis.
At the same time
the Thomas-Wigner rotation angle
increases to $+360^\circ$
as $\gamma\rightarrow\infty$
(see the discussion in section~\ref{se:discuss} below).

Since the accelerated object is Born-rigid,
the trajectories of all grid vertices~$G$
are uniquely determined
once the trajectory of the reference point~$R$ is known
\citep{herglotz09,noether10,eriksen82}.
Following the discussion in section~\ref{se:uniformaccmot}
the position and coordinate time of a random vertex~$G$,
in the frame comoving with $R$ at the beginning
of the corresponding acceleration phase,
follows from eqns.~\ref{eq:postimattauB}.
The resulting trajectories are discussed
in the next section.

\section{Visualization}
\label{se:visual}

The trajectory of the reference point~$R$
within the $xy$-plane
for a boost velocity of $\beta = 1/2$,
corresponding to $\gamma = 2/\sqrt{3} \approx 1.15$,
is displayed in fig.~\ref{fg:trjbstrefptfr1and6}.
It is plotted
in the laboratory frame
(frame~$[1]$) as black solid line.
The same trajectory
as it appears to an observer in frame~$[6]$
is marked in grey.
The two frames are stationary with respect to other,
but rotated by a Thomas-Wigner angle
of about~$14.4^\circ$.
In addition,
dots show the spatial component
of the four switchover events
$\text{B}$, $\text{C}$, $\text{D}$ and $\text{E}$
in the two frames.
As required by
assumption~[\ref{as:posvelstartend}]
the start and final positions,
corresponding to the events $\text{A}$ and $\text{F}$,
coincide.

\begin{figure}
\includegraphics{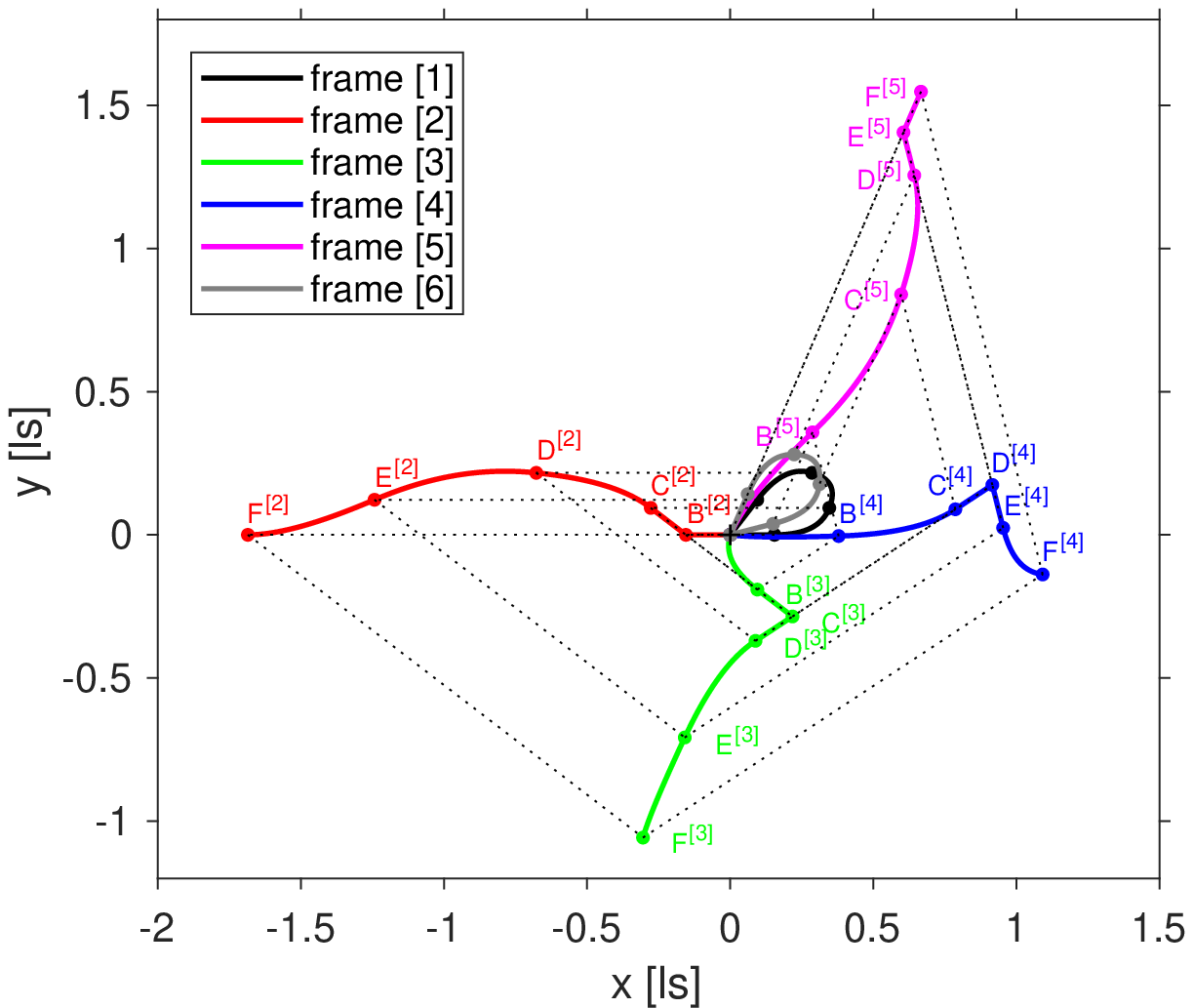}
\caption{
\label{fg:trjboostrefpt}
Trajectories of the reference point~$R$
as seen from the six reference frames
$[1],[2],\ldots, [6]$.
The switchover points are marked by $X_i^{[k]}$
with $X = \text{A},\ldots,\text{F}$.
Corresponding switchover points
are connected
by dashed lines.
The Lorentz factor is
$\gamma = 2/\sqrt{3} \approx 1.15$.
}
\end{figure}

Fig.~\ref{fg:trjboostrefpt} shows the same trajectory
as fig.~\ref{fg:trjbstrefptfr1and6}.
In addition,
$R$'s trajectories
as recorded by observers in the
frames~$[2],\ldots,[5]$
are plotted as well (solid coloured lines).
Corresponding switchover events are connected
by dashed lines.
At $\text{B}^{[2]}$, $\text{C}^{[3]}$, $\text{D}^{[4]}$ and $\text{E}^{[5]}$
(and of course at the start event $\text{A}^{[1,6]}$
and destination event $\text{F}^{[1,6]}$)
the reference point~$R$ slows down and/or accelerates
from zero velocity
producing a kink in the trajectory.
In all other cases
the tangent vectors of the trajectories,
i.e.\ the velocity vectors
are continuous at the switchover points.

\begin{figure}
\includegraphics{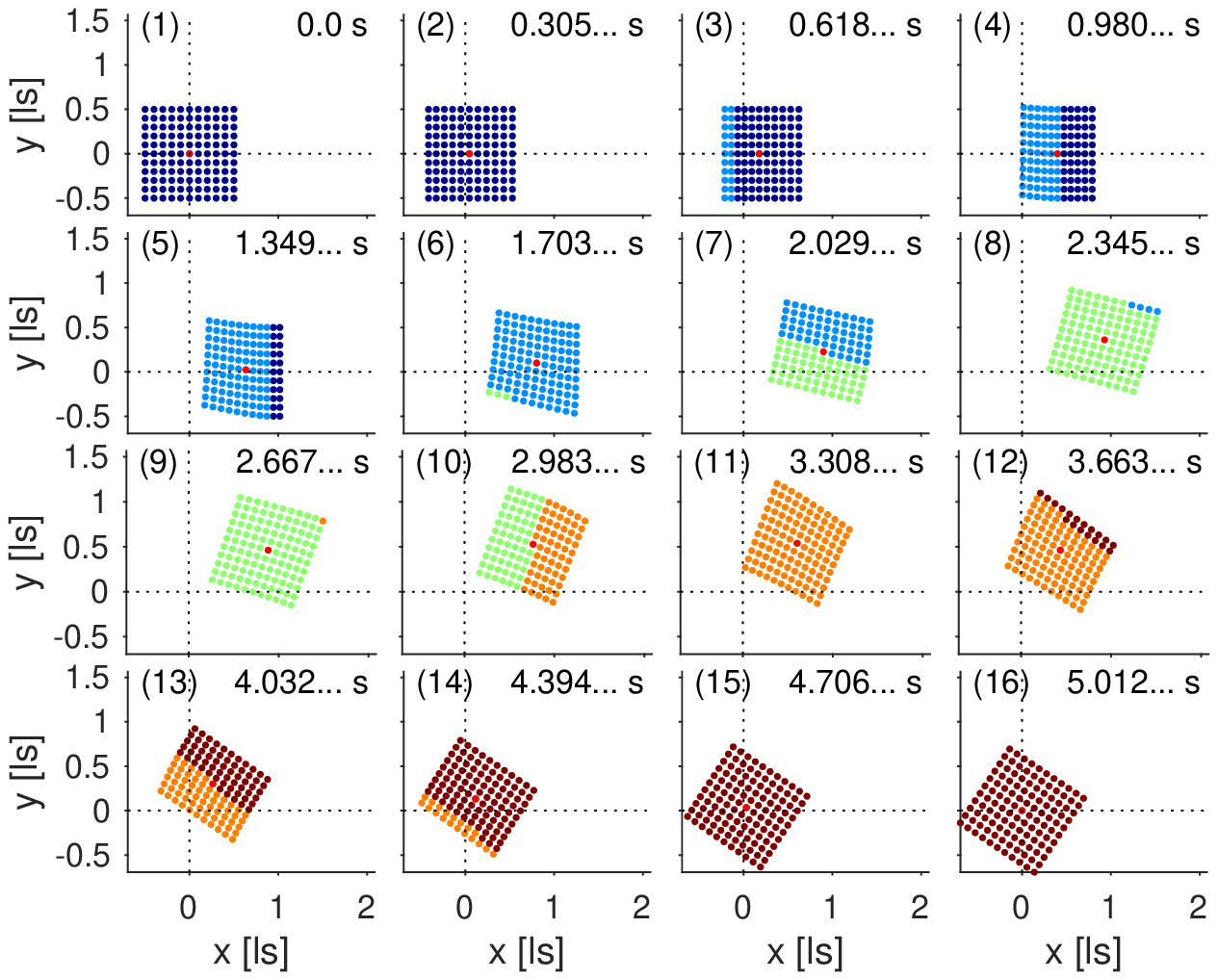}
\caption{
\label{fg:trjboostobj}
A series of grid positions
as seen in the laboratory frame.
The boost velocity is taken to be $\beta = 0.7$,
the resulting Thomas-Wigner rotation angle
amounts to about~$33.7^\circ$.
Coordinate time is displayed
in the top right corner of each panel.
The five boost phases
are distinguished by colour.
It is evident that
switchovers between boosts
do not occur simultaneously
in the laboratory frame.
The reference point
(marked in red)
moves along its trajectory
counterclockwise,
whereas the grid
Thomas-Wigner rotates clockwise.
For details see text.
}
\end{figure}

With
eqns.~\ref{eq:solvepol4deg} and
\ref{eq:t23fromt12sel}
all necessary ingredients
to visualize the relativistic motion
of a Born-rigid object are available.
In fig.~\ref{fg:trjboostobj}
the object is modelled
as a square-shaped grid of~$11\times11$ points,
arranged around the reference point~$R$.
The object uniformly accelerates
in the $xy$-plane
changing the boost directions
by the four angles
$\zeta_{1,2}$ (as measured in frame~$[2]$),
$\zeta_{2,3}$ (measured in frame~$[3]$),
$\zeta_{2,3}$ (measured in frame~$[4]$)
and finally $\zeta_{1,2}$ (measured in frame~$[5]$).
The vertices' colour code
indicates the corresponding boost section.
The 16~panels depict the grid positions
in the laboratory frame (frame~$[1]$)
for specific values of coordinate time
displayed in the top right.

To improve the visual impression
the magnitude of the Thomas-Wigner rotation
in fig.~\ref{fg:trjboostobj}
is enlarged
by increasing the boost velocity
from $\beta = 0.5$
in figs.~\ref{fg:trjbstrefptfr1and6} and \ref{fg:trjboostrefpt}
to $\beta = 0.7$
corresponding to $\gamma \approx 1.4$.
Despite appearance
the grid~$\mathcal{G}$ is Born-rigid;
in~$R$'s comoving inertial frame
the grid maintains its original square shape.
In the laboratory frame, however,
$\mathcal{G}$ appears compressed,
when it starts to accelerate or decelerate
and sheared,
when one part of~$\mathcal{G}$
has not yet finished boost~$k$,
but the remaining part of~$\mathcal{G}$
already has transitioned to the next boost section~$k+1$.
This feature is clearly evident
from panels~(4), (7), (10) or (13)
in fig.~\ref{fg:trjboostobj}
with
the occurrence of two colours
indicating two boost sections taking effect
at the same epoch of coordinate time.
We note, however, that the switchover events
occur simultaneously for all grid points
in $R$'s comoving frame.
The non-uniform colourings
illustrate the asynchronism of the switchovers
in the laboratory frame
and
thereby the relationship
between Thomas-Wigner rotations
and the non-existence of absolute simultaneity.

\section{Discussion}
\label{se:discuss}

In this final section
the Thomas-Wigner rotation angle
is calculated from the known boost angles
$\zeta_{1,2}(\gamma)$
and
$\zeta_{2,3}(\gamma)$
(eqn.~\ref{eq:zeta1fromgamma}).
Second,
the question of size limits of Born-rigid objects,
Thomas-Wigner-rotated by a series of boosts,
is addressed.

\subsection{Derivation of Thomas-Wigner angle}
\label{su:derivethomaswigner}

\begin{figure}
\includegraphics{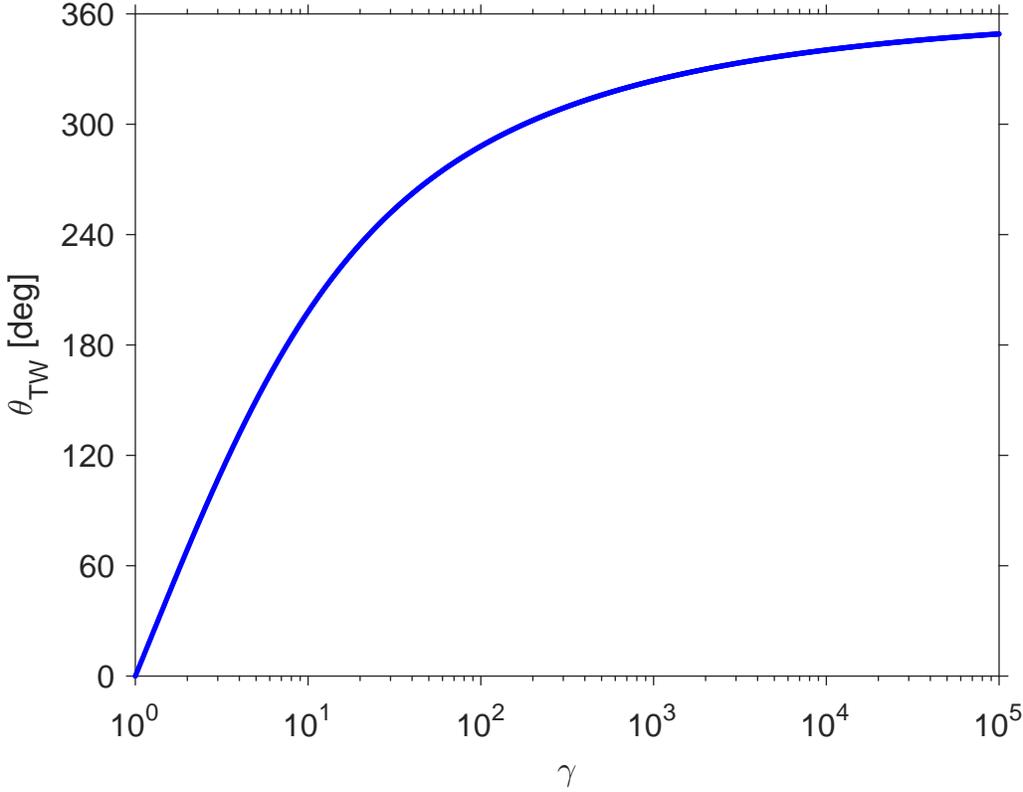}
\caption{
\label{fg:thomaswigneranglevsgamma}
Thomas-Wigner rotation angle as a function of $\gamma$.
For clarity the angle
has been unwrapped to the range $[0^\circ, +360^\circ]$.
}
\end{figure}

From the preceding sections follows
a straightforward calculation
of the Thomas-Wigner angle
as a function of Lorentz factor~$\gamma$.
Assumption~[\ref{as:posvelstartend}] implies
that the sequence of the five Lorentz transformations
$[6] \rightarrow [5] \rightarrow \ldots \rightarrow [1]$
is constructed such that frame $[6]$
is stationary with respect to frame $[1]$
and their spatial origins coincide.
I.e.\ the combined transformation
reduces to an exclusively spatial rotation
and the corresponding Lorentz matrix
can be written as
\begin{eqnarray}
\label{eq:thomaswignerrotmatrix}
&& \Lambda\left(\gamma,-\hat{e}_1\right)    \cdot
  \, \Lambda\left(\gamma,-\hat{e}_2\right) \cdot
  \, \Lambda\left(\gamma,-\hat{e}_3\right) \cdot
  \, \Lambda\left(\gamma,-\hat{e}_4\right) \cdot
  \, \Lambda\left(\gamma,-\hat{e}_5\right)
  \\
&=&
\begin{pmatrix}
    1
    & 0
    & 0
    & 0 \\
    0
    & R_{1,1}
    & R_{1,2}
    & R_{1,3} \\
    0
    & R_{2,1}
    & R_{2,2}
    & R_{2,3} \\
    0
    & R_{3,1}
    & R_{3,2}
    & R_{3,3}
\end{pmatrix}
\quad .
\nonumber
\end{eqnarray}
Since the rotation is confined to the $xy$-plane,
$R_{3,i} = 0 = R_{i,3}$ with $i=1,2,3$.
The remaining elements
\begin{eqnarray}
\label{eq:thomaswignerrotmatelem}
R_{1,1}(\gamma)
  &=& R_{2,2}(\gamma)
  \equiv \cos(\theta_{TW}(\gamma))
\\
R_{1,2}(\gamma)
  &=& -R_{2,1}(\gamma)
  \equiv \sin(\theta_{TW}(\gamma))
\nonumber
\end{eqnarray}
yield the Thomas-Wigner rotation angle
\begin{eqnarray}
\label{eq:thowigrotangle}
\tilde{\theta}_{TW}(\gamma)
  \equiv
\text{atan2}( R_{2,1}(\gamma), R_{1,1}(\gamma))
\\
\theta_{TW}(\gamma)
  \equiv
\left\{
\begin{array}
{c@{\quad:\quad}l}
\tilde{\theta}_{TW}(\gamma)
  &
  \tilde{\theta}_{TW}(\gamma) \ge 0
  \\
\tilde{\theta}_{TW}(\gamma) + 2\,\pi
  &
  \tilde{\theta}_{TW}(\gamma) < 0
\end{array}
\right.
\nonumber
\end{eqnarray}
with $\text{atan2}(\cdot,\cdot)$ denoting
the four-quadrant inverse tangent.
The definition of $\theta_{TW}(\gamma)$
in eqn.~\ref{eq:thowigrotangle} ensures
that angles exceeding $+180^\circ$
are unwrapped and mapped into the interval $[0^\circ,+360^\circ]$
(see fig.~\ref{fg:thomaswigneranglevsgamma}).

Using eqn.~\ref{eq:thomaswignerrotmatrix}
the rotation matrix elements
$R_{1,1}$ and $R_{2,1}$,
expressed in terms of
$T_{12}$ and $T_{23}$
(eqn.~\ref{eq:halfangleparm}),
are (see~\ref{se:appendix})
\begin{eqnarray}
\label{eq:thwirot11}
R_{1,1}(\gamma)
&=&
  -1 + \frac{\left(\gamma + 1\right)}
            {\left((T_{12})^2 + 1\right)^4
             \left((T_{23})^2 + 1\right)^4}
  \\
&& \times
  \bigg((T_{12})^4 \, (T_{23})^4
  + 2 \, (T_{12})^4 \, (T_{23})^2
  + (T_{12})^4
  \nonumber
  \\
&&
  - 4 \, (T_{12})^3 \, (T_{23})^3 \, \gamma
  + 4 \, (T_{12})^3 \, (T_{23})^3
  - 4 \, (T_{12})^3 \, T_{23} \, \gamma
  \nonumber
  \\
&&
  + 4 \, (T_{12})^3 \, T_{23}
  - 2 \, (T_{12})^2 \, (T_{23})^4 \, \gamma
  + 4 \, (T_{12})^2 \, (T_{23})^4
  \nonumber
  \\
&&
  + 4 \, (T_{12})^2 \, (T_{23})^2 \, \gamma^2
  + 8 \, (T_{12})^2 \, (T_{23})^2 \, \gamma
  - 8 \, (T_{12})^2 \, (T_{23})^2
  \nonumber
  \\
&&
  - 4 \, (T_{12})^2 \, \gamma^2
  + 10 \, (T_{12})^2 \, \gamma
  - 4 \, (T_{12})^2
  \nonumber
  \\
&&
  + 12 \, T_{12} \, (T_{23})^3 \, \gamma
  - 12 \, T_{12} \, (T_{23})^3
  - 16 \, T_{12} \, T_{23} \, \gamma^2
  \nonumber
  \\
&&
  + 12 \, T_{12} \, T_{23} \, \gamma
  + 4 \, T_{12} \, T_{23}
  + 2 \, (T_{23})^4 \, \gamma
  \nonumber
  \\
&&
  - (T_{23})^4
  - 4 \, (T_{23})^2 \, \gamma^2
  + 6 \, (T_{23})^2
  + 4 \, \gamma^2
  - 2 \, \gamma - 1\bigg)^2
  \nonumber
\end{eqnarray}
and
\begin{eqnarray}
\label{eq:thwirot21}
R_{2,1}(\gamma)
  &=&
  \frac{4 \left( \gamma - 1\right) \,
  \left(\gamma + 1\right)}{\left((T_{12})^2 + 1\right)^4
  \left((T_{23})^2 + 1\right)^4}
  \\
&& \times
  \bigg((T_{12})^3 \, (T_{23})^2
  + (T_{12})^3
  + 3 \, (T_{12})^2 \, (T_{23})^3
  \nonumber
  \\
&&
  - 2 \, (T_{12})^2 \, T_{23} \, \gamma
  + (T_{12})^2 \, T_{23}
  + T_{12} \, (T_{23})^4
  \nonumber
  \\
&&
  - 2 \, T_{12} \, (T_{23})^2 \, \gamma
  - 3 \, T_{12} \, (T_{23})^2
  + 2 \, T_{12} \, \gamma
  \nonumber
  \\
&&
  - (T_{23})^3
  + 2 \, T_{23} \, \gamma
  + T_{23}\bigg)
  \nonumber
  \\
&& \times
  \bigg((T_{12})^4 \, (T_{23})^4
  + 2 \, (T_{12})^4 \, (T_{23})^2
  + (T_{12})^4
  \nonumber
  \\
&&
  - 4 \, (T_{12})^3 \, (T_{23})^3 \, \gamma
  + 4 \, (T_{12})^3 \, (T_{23})^3
  - 4 \, (T_{12})^3 \, T_{23} \gamma
  \nonumber
  \\
&&
  + 4 \, (T_{12})^3 \, T_{23}
  - 2 \, (T_{12})^2 \, (T_{23})^4 \gamma
  + 4 \, (T_{12})^2 \, (T_{23})^4
  \nonumber
  \\
&&
  + 4 \, (T_{12})^2 \, (T_{23})^2 \gamma^2
  + 8 \, (T_{12})^2 \, (T_{23})^2 \gamma
  - 8 \, (T_{12})^2 \, (T_{23})^2
  \nonumber
  \\
&&
  - 4 \, (T_{12})^2 \, \gamma^2
  + 10 \, (T_{12})^2 \, \gamma
  - 4 \, (T_{12})^2
  \nonumber
  \\
&&
  + 12 \, T_{12} \, (T_{23})^3 \, \gamma
  - 12 \, T_{12} \, (T_{23})^3
  - 16 \, T_{12} \, T_{23} \, \gamma^2
  \nonumber
  \\
&&
  + 12 \, T_{12} \, T_{23} \, \gamma
  + 4 \, T_{12} \, T_{23}
  + 2 \, (T_{23})^4 \, \gamma
  \nonumber
  \\
&&
  - (T_{23})^4
  - 4 \, (T_{23})^2 \, \gamma^2
  + 6 (T_{23})^2 + 4 \gamma^2
  - 2 \gamma - 1\bigg)
  \nonumber
\end{eqnarray}
with
$T_{12} = T_{12}(\gamma)$ and $T_{23} = T_{23}(\gamma)$
given by eqns.~\ref{eq:solvepol4deg}
and
\ref{eq:t23fromt12sel},
respectively.

The resulting angle $\theta_{TW}(\gamma)$
as a function of $\gamma$
is plotted
in fig.~\ref{fg:thomaswigneranglevsgamma}.
The plot suggests that
$\theta_{TW} \rightarrow +360^\circ$
as $\gamma \rightarrow \infty$.
As already mentioned in subsection~\ref{su:fiveboosts}
(see fig.~\ref{fg:normboostdirs})
the boost angles
$\zeta_{1,2} \rightarrow +180^\circ$
and
$\zeta_{2,3} \rightarrow +180^\circ$
in the relativistic limit
$\gamma \rightarrow \infty$.
Notwithstanding
that the $R$'s trajectory
reduces to an one-dimensional motion
along the $x$-axis
as
$\gamma \rightarrow \infty$,
the grid's Thomas-Wigner rotation angle
approaches a full revolution of $+360^\circ$
in the laboratory frame.

\subsection{Event horizons}
\label{su:eventhorizon}

As illustrated by fig.~\ref{fg:trjboostobj}
the Born-rigid object~$\mathcal{G}$
rotates in the $xy$-plane.
Clearly,
$\mathcal{G}$'s spatial extent
in the $x$- and $y$-directions
has to be bounded
by a maximum distance
from the reference point~$R$
on the order of
$\Delta t / \theta_{TW}$
\citep{born09}.
This boundary,
which prevents paradoxical
faster-than-light translations
of sufficiently distant vertices,
is produced by event horizons
associated with the accelerations
in the five boost sections.

\begin{figure}
\includegraphics{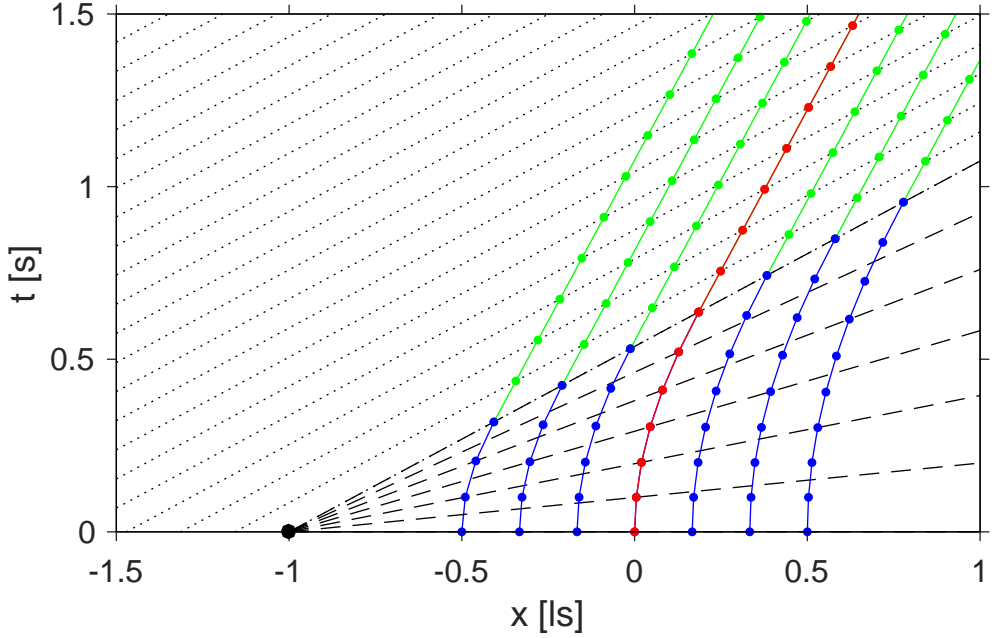}
\caption{
\label{fg:accgrid1dimevhrz}
Spacetime diagram
of a one-dimensional grid consisting of seven points.
The grid accelerates towards
the positive $x$-direction.
The trajectories are marked in blue,
the mid point is taken as the reference~$R$
and its worldline is colored in red.
Dots indicate the lapse of 0.1~s in proper time.
After 0.6~s have passed on $R$'s proper time close
the acceleration stops
and the points move with constant velocity
(green lines).
Dashed lines connect spacetime
points of $R$'s comoving frame.
The grey area indicates
the inaccessible spacetime
beyond the event horizon.
}
\end{figure}

Fig.~\ref{fg:accgrid1dimevhrz}
exemplifies
the formation of an event horizon
for an accelerated object
in $1+1$ (one time and one space) dimensions
\citep[see e.g.][]{desloge87,eriksen82,hamilton78,semay06}.
Here, the Born-rigid object is assumed
to be one-dimensional and
to consist of seven equidistant grid points.
Each point accelerates for a finite time period
towards the positive $x$-direction
(blue worldlines);
the reference point~$R$,
marked in red,
accelerates with
$\alpha_R \equiv 1~\text{ls}/\text{s}^2$.
Contrary to the simulations
discussed
in fig.~\ref{fg:trjboostobj} above,
in the present case
the acceleration phase is not followed
immediately by another boost.
Rather, the object continues
to move with constant velocity
after the accelerating force has been switched off
(green worldlines in fig.~\ref{fg:accgrid1dimevhrz}).
The completion of the acceleration phase
is synchronous in $R$'s
instantaneous comoving frame
(dashed-dotted line)
and asynchronous in the laboratory frame.
Fig.~\ref{fg:accgrid1dimevhrz} also illustrates
that for an uniform acceleration
the event horizon (black dot)
is stationary with respect to the laboratory frame.

In this simulation
each vertex is assumed to be equipped with an ideal clock
ticking at a proper frequency of 10~Hz,
the corresponding ticks are marked by dots;
the boost phase lasts for 0.6~s
on $R$'s clock.
The clocks
of the left-most (trailing)
and right-most (leading) vertex
measure (proper time) boost durations
of 0.3~s and 0.9~s, respectively.
Thus, with respect to the
instantaneous comoving frames
(dashed  lines)
the vertex clocks run at different rates
(see eqn.~\ref{eq:tautfromtaur}).
The trailing clocks tick slower,
the leading clocks faster
than the reference clock at~$R$.
From eqns.~\ref{eq:alphatfromalphar} and \ref{eq:tautfromtaur}
it follows
that the proper time variations
are compensated by
corresponding changes in proper acceleration
experienced by the seven vertices.
For the parameters used in fig.~\ref{fg:accgrid1dimevhrz}
the accelerations of the trailing
and leading vertex
are $2\,\alpha_R$ and $2\,\alpha_R/3$, respectively.

The spatial components of the inertial reference frames,
comoving with~$R$,
are plotted in fig.~\ref{fg:accgrid1dimevhrz}
as well.
During the acceleration-free period
following the boost phase
the grid moves with constant velocity
and
the equal-time slices of
the corresponding comoving frames
(dotted lines)
are oriented parallel to other.
During the boost phase,
however,
the lines intersect
and eqn.~\ref{eq:postimattauB} entails
that the equal-time slices
of the comoving frames
all intersect
in one spacetime point,
the event horizon
$x_{H} \equiv -1~\text{ls}$
(black dot at $x = -1$~ls and $t = 0$~s
in fig.~\ref{fg:accgrid1dimevhrz}).
A clock placed at $x_{H}$ does not tick,
time is frozen at this spacetime point
and
eqns.~\ref{eq:alphatfromalphar} and \ref{eq:tautfromtaur}
imply that the corresponding proper acceleration diverges.
Clearly
a physical object
accelerating towards positive~$x$
in fig.~\ref{fg:accgrid1dimevhrz}
cannot extend beyond
this boundary at $x_{H}$.
If the grid
in fig.~\ref{fg:accgrid1dimevhrz}
is regarded as realization
of an accelerating coordinate system,
this frame
is bounded in the spatial dimension
and ends at the coordinate value $x_H$.
However,
as soon as the grid's acceleration stops,
the event horizon disappears
and
coordinates $x < x_H$
are permissible.
We note, that the event horizon
in fig.~\ref{fg:accgrid1dimevhrz}
is a zero-dimensional object,
a point in $1+1$-dimensional spacetime
considered here.
The horizon is frozen in time
and exists only for the instant $t=0$.

Generalizing this result
we find that the five boosts
described
in subsection~\ref{su:fiveboosts}
and
depicted in fig.~\ref{fg:trjbstrefptfr1and6}
induce five event horizons
in various orientations.
It turns out
that the accelerated object~$\mathcal{G}$
is bounded
by these horizons
in all directions
within the $xy$-plane.
They limit $\mathcal{G}$'s maximum size
\citep{born09}
and thereby assure that all of its vertices
obey the special relativistic speed limit
\citep{einstein05}.

\section{Conclusions}
\label{se:conclusion}

It is well known that pure Lorentz transformations
do not form a group in the mathematical sense,
since the composition of two transformations
in general is not a pure Lorentz transformation again,
but involves the Thomas-Wigner spatial rotation.
The rotation is illustrated
by uniformly accelerating
a Born-rigid object,
which is assumed to consist
of a finite number of vertices,
repeatedly,
such that the object's reference point
returns to its start location.
It turns out that at leasts five boosts
are necessary,
provided,
first, the (proper time) duration
and the magnitude of the proper acceleration
within each boost is the same
and, second,
the object's motion is restricted
to the $xy$-plane.
Analytic expressions are derived for
the angles between adjacent boost directions.

The visualizations illustrate the relationship
between Thomas-Wigner rotations
and the relativity of simultaneity.
The transition from one boost section
to the next
occurs synchronously
in the instantaneous comoving frame.
In the laboratory frame, however,
the trailing vertices
conclude the current boost phase first
and switch to the next one,
which in general involves
a direction change.
In the laboratory frame
the accelerated object not only
contracts and expands
along its direction of propagation,
but also exhibits a shearing motion
during the switchover phases.
The simulations illustrate clearly
that the aggregation of these shearing
contributions
finally adds up to the
Thomas-Wigner rotation.

Accelerated motions induce event horizons,
which no part of a physical, Born-rigid object
may overstep.
Thus,
the object's size is limited
to a finite volume or area
(if its motion is restricted to two spatial dimension)
and
Thomas-Wigner rotations by construction
observe the special relativistic speed limit.

\section{Auxiliary material}

An MPEG-4 movie showing the Thomas-Wigner rotation
of a grid object
is available at the URL \url{www.gbeyerle.de/twr}.
In addition, the \textsc{Matlab} source code
used to create fig.~\ref{fg:trjboostobj}
and the ``SymPy''
script file
discussed in~\ref{se:appendix}
can be downloaded from the same site.

\ack
Some calculations described in this paper
were performed with the help
of the computer algebra system ``SymPy''
\citep{joyner12},
which is available for download from
\url{www.sympy.org}.
``SymPy'' is licensed
under the General Public License;
for more information see
\url{www.gnu.org/licenses/gpl.html}\;.
Trademarks are the property of their respective owners.

\appendix
\section{}
\label{se:appendix}

A number of equations in this paper
were derived
using the computer algebra system ``SymPy''
\citep{joyner12}.
The corresponding ``SymPy'' source code files
\texttt{vtwr3bst.py}
(three boost case, see subsection~\ref{su:threeboosts})
and
\texttt{vtwr5bst.py}
(five boost case, see subsection~\ref{su:fiveboosts})
are available for download at \url{www.gbeyerle.de/twr}.
These scripts process
eqns.~\ref{eq:threeboostpos}, \ref{eq:threeboostvel},
\ref{eq:fiveboostpos} and \ref{eq:fiveboostvel}
and derive the results given in
eqns.~\ref{eq:threebooststvalsol},
\ref{eq:threeboostspossol},
\ref{eq:t23fromt12},
\ref{eq:t12t23first},
\ref{eq:t12t23second},
\ref{eq:thwirot11} and
\ref{eq:thwirot21}.
The following
offers a few explanatory comments.

First,
we address the case of three boosts
(``SymPy'' script \texttt{vtwr3bst.py})
and the derivation of
eqn.~\ref{eq:threebooststvalsol}.
The corresponding boost vectors
in the $xy$-plane
$\hat{e}^{(3B)}_i$ with $i=1,2,3$
are taken to be
\begin{eqnarray}
\label{eq:enrm2bst3}
\hat{e}^{(3B)}_1
\equiv
\begin{pmatrix}
  C_a \\
  S_a
\end{pmatrix}
=
\frac{1}{1+(T_{12})^2}
\begin{pmatrix}
  1-(T_{12})^2 \\
  2\,T_{12}
\end{pmatrix}
\\
\hat{e}^{(3B)}_2
\equiv
\begin{pmatrix}
  1 \\
  0
\end{pmatrix}
\nonumber \\
\hat{e}^{(3B)}_3
\equiv
\begin{pmatrix}
  C_a \\
  -S_a
\end{pmatrix}
=
\frac{1}{1+(T_{12})^2}
\begin{pmatrix}
  1-(T_{12})^2 \\
  -2\,T_{12}
\end{pmatrix}
\nonumber
\end{eqnarray}
with
\begin{eqnarray}
\label{eq:defs1c1}
S_a &\equiv& \sin(\zeta_{1,2})
\qquad
C_a \equiv \cos(\zeta_{1,2})
\end{eqnarray}
in terms of the direction angle~$\zeta_{1,2}$
and the half-angle approximation
(eqn.~\ref{eq:halfanglesincos}).
Here,
the $z$-coordinate has been omitted
since the trajectory is restricted to the $xy$-plane
and
$\zeta_{2,3} = \zeta_{1,2}$
from time reversal symmetry is being used.
Inserting the corresponding Lorentz transformation matrices
(eqn.~\ref{eq:lotra1to2})
into eqn.~\ref{eq:threeboostvel}
and selecting the time component
yields
\begin{eqnarray}
\label{eq:eval3bvel}
(\gamma - 1)\,
  \frac{((T_{12})^2 - 2\,\gamma - 1)^2}
       {((T_{12})^2 + 1)^2} = 0
\end{eqnarray}
which reduces to eqn.~\ref{eq:threebooststvalsol}
if the trivial solution $\gamma=1$ is ignored.

The case of five boosts
(``SymPy'' script \texttt{vtwr5bst.py})
and the derivation of
eqn.~\ref{eq:t12t23first}
is addressed next.
In analogy to eqn.~\ref{eq:enrm2bst3}
we define
\begin{eqnarray}
\label{eq:enrm2bst5}
\hat{e}^{(5B)}_1
\equiv
\begin{pmatrix}
  C_x \\
  S_x
\end{pmatrix}
=
\begin{pmatrix}
  C_{12}\,C_{23} - S_{12}\,S_{23} \\
  -S_{12}\,C_{23} - C_{12}\,S_{23}
\end{pmatrix}
\\
\hat{e}^{(5B)}_2
\equiv
\begin{pmatrix}
  C_y \\
  S_y
\end{pmatrix}
=
\begin{pmatrix}
  C_{23} \\
  -S_{23}
\end{pmatrix}
=
\frac{1}{1+(T_{23})^2}
\begin{pmatrix}
  1-(T_{23})^2 \\
  -2\,T_{23}
\end{pmatrix}
\nonumber
\\
\hat{e}^{(5B)}_3
\equiv
\begin{pmatrix}
  1 \\
  0
\end{pmatrix}
\nonumber \\
\hat{e}^{(5B)}_4
\equiv
\begin{pmatrix}
  C_y \\
  -S_y
\end{pmatrix}
=
\begin{pmatrix}
  C_{23} \\
  S_{23}
\end{pmatrix}
=
\frac{1}{1+(T_{23})^2}
\begin{pmatrix}
  1-(T_{23})^2 \\
  2\,T_{23}
\end{pmatrix}
\nonumber
\\
\hat{e}^{(5B)}_5
\equiv
\begin{pmatrix}
  C_x \\
  -S_x
\end{pmatrix}
=
\begin{pmatrix}
  C_{12}\,C_{23} - S_{12}\,S_{23} \\
  S_{12}\,C_{23} + C_{12}\,S_{23}
\end{pmatrix}
\nonumber
\end{eqnarray}
with eqn.~\ref{eq:defs1c1} and
\begin{eqnarray}
\label{eq:defs2c2}
S_{12} &\equiv& \sin(\zeta_{1,2})
\qquad
C_{12} \equiv \cos(\zeta_{1,2})
\\
S_{23} &\equiv& \sin(\zeta_{2,3})
\qquad
C_{23} \equiv \cos(\zeta_{2,3})
\quad .
\nonumber
\end{eqnarray}
The corresponding Lorentz transformation matrices
are too unwieldy to reproduce them here.
``SymPy'' script \texttt{vtwr5bst.py}
calculates them, their products
in terms of $T_{12}$ and $T_{23}$
and
inserts them into eqn.~\ref{eq:fiveboostvel}.
Selecting the time component
yields the equation
\begin{eqnarray}
\label{eq:eqt12t23for5boost}
\frac{\gamma - 1}
     {((T_{12})^2 + 1)^2\,((T_{23})^2 + 1)^2}
     \\
\times
  \left((T_{12})^2\,(T_{23})^2
            + (T_{12})^2
            - 4\,T_{12}\,T_{23}\,\gamma
            \right.
\nonumber
\\
  \left. \qquad
         - 4\,T_{12}\,T_{23}
         - 2\,(T_{23})^2\,\gamma
         - (T_{23})^2 + 4\,\gamma^2
         + 2\,\gamma - 1\right)^2
        = 0
\quad .
\nonumber
\end{eqnarray}
We exclude the trivial solution $\gamma=1$
and restrict ourselves
to real values of $T_{12}$ and $T_{23}$;
eqn.~\ref{eq:eqt12t23for5boost}
then leads to
eqn.~\ref{eq:t12dott23},
a second order polynomial with respect
to $T_{23}$.
The two solutions are given in
eqn.~\ref{eq:t23fromt12}.

Next insert $T_{23} = T_{23}(T_{12},\gamma)$
in eqn.~\ref{eq:fiveboostpos}.
Since its time component
involves the travel time of $R$ along its closed trajectory
as an additional unknown
and the $z$-coordinate vanishes by construction,
we focus in the following on the $x$- and $y$-components
of eqn.~\ref{eq:fiveboostpos}.
The script \texttt{vtwr5bst.py} shows that
the result for the $y$-component
of the four-vector equation
$\vec{P}_F^{[6]} = 0$
can be expressed in the form
\begin{eqnarray}
\label{eq:}
-8\,((T_{12})^2 + 1)\,(\gamma + 1)^2
  \, \frac{X_1(T_{12},\gamma) + X_2(T_{12},\gamma) \, \sqrt{X_3(T_{12},\gamma)}}
          {((T_{12})^2 - 2\,\gamma - 1)^6} = 0
\end{eqnarray}
Here,
$X_i(T_{12},\gamma)$ with $i=1,2,3$
are polynomials in $T_{12}$.

For real $T_{12}$ and $\gamma \ge 1$
the numerator has to equate to zero.
Moving the term involving the square root to the RHS
and squaring both sides
yields
\begin{eqnarray}
\left(X_1(T_{12},\gamma)\right)^2 -
  \left(X_2(T_{12},\gamma)\right)^2 \, X_3(T_{12},\gamma) = 0
  \quad .
\end{eqnarray}
Evaluation of this expression (see script \texttt{vtwr5bst.py})
leads to the product of two polynomial expression,
each of which is fourth order with respect to $(T_{12})^2$
(see eqn.~\ref{eq:t12t23first} and \ref{eq:t12t23second}).

Repeating these calculations
for the $x$-component of the four-vector equation
$\vec{P}_F^{[6]} = 0$
leads also to the product of two polynomial expression,
one of which is identical to
eqn.~\ref{eq:t12t23first}.
Thus,
we have found a solution to
eqn.~\ref{eq:posinit5boost}.

The Thomas-Wigner angle $\theta_{TW}$
(eqn.~\ref{eq:thowigrotangle})
follows from the Lorentz matrix
relating frame~$[1]$ to frame $[6]$
(eqn.~\ref{eq:thomaswignerrotmatrix}).
Script \texttt{vtwr5bst.py}
evaluates the matrix elements
$R_{1,1}$ and $R_{2,1}$
in terms of the variable $T_{12}$ and $T_{23}$.
Again the expressions are unwieldy,
but the derivation is straightforward.

\end{document}